\documentstyle[12pt,eqsecnum,multicol,epsfig,aps,prl,array]{revtex}
\newcommand{\bleq}{\ifpreprintsty
		   \else
		   \end{multicols}\widetext \vspace*{-3.5ex}{\tiny
		   
		\noindent\begin{tabular}[t]{c|}
		   \parbox{0.493\hsize}{~} \\ \hline \end{tabular}}
				      \fi}
\newcommand{\eleq}{\ifpreprintsty
		   \else
		   {\tiny\hspace*{\fill}\begin{tabular}[t]{|c}\hline
		    \parbox{0.49\hsize}{~} \\
		    \end{tabular}}\vspace*{-2.5ex}\begin{multicols}{2}
		    \narrowtext
		    \fi}
\newcommand{\bcols}{\ifpreprintsty\else\begin{multicols}{2} 
	\narrowtext\fi}
\newcommand{\ecols}{\ifpreprintsty\else\end{multicols}\fi}

\draft
\widetext
\begin{document}
\title{Direct evidence of rigidity loss and self-organisation in silicate glasses} 
\author{Y. Vaills$^1$, T. Qu$^2$, M. Micoulaut$^3$, F. Chaimbault$^1$
and P. Boolchand$^2$}
\address{\ \\
$^1$ Université d'Orléans  45067 Orléans Cedex 02 and\\ 
Centre de Recherches sur les Matériaux à Haute Température, 
1D avenue de la recherche scientifique,\\
 45071 Orléans Cedex 02, France\\
$^2$ Department of Electrical and Computer Engineering and Computer Science, 
University of Cincinnati, OH 45221-0030\\
$^3$Laboratoire de Physique Théorique des Liquides, Université Pierre et 
Marie Curie, Boite 121\\
4, Place Jussieu, 75252 Paris Cedex 05, France}

\date{\today}
\maketitle
$$ $$
\begin{abstract}
The Brillouin elastic free energy change $\Delta\Phi$ between thermally  
annealed and quenched $(Na_2O)_x(SiO_2)_{1-x}$ glasses is found to decrease 
linearly  at  $x > 0.23$ ( floppy phase), and to nearly vanish at $x  < 0.18$
 (stressed- rigid  phase). The observed $\Delta \Phi(x)$ variation closely 
parallels the mean-field floppy mode fraction $f(x)$ in random networks, 
and fixes the two (floppy, stressed-rigid) elastic phases. 
In calorimetric measurements, the non-reversing enthalpy near $T_g$ is found 
to be large at $x < 0.18$ and at $x > 0.23$ , but to nearly vanish in the 
$0.18 < x < 0.23$ range, suggesting existence of an intermediate phase 
between the floppy and stressed-rigid phases.
\par
{Pacs:} 61.43Fs, 63.50.+x
\end{abstract}
\newpage
Silicate melts and glasses are important geophysical\cite{1}, 
optoelectronic\cite{2} and microelectronic materials\cite{3}, and find 
applications as window glass materials, optical fibers and thin-film 
gate dielectrics\cite{3}. Functionality of materials often derives from their 
structures at different length scales. At a basic level, the molecular 
structure of sodium silicate $(Na_2O)_x(SiO_2)_{1-x}$ glasses consist of a network 
of $Si(O_{1/2})_4$ tetrahedra in which addition of sodium oxide produces 
$Si(O_{1/2})_{4-m} O_m Na_m$  (or $Q^{4-m}$ in NMR notation) local
units having $m  = 0,1,2$ and 
$3$ non-bridging oxygen (NBO) sites attached to $Na^+$ ions. Addition of  a 
few (10) mole percent of $Na_2O$ lowers\cite{4}  
the glass transition temperature, 
$T_g$, of the base ($SiO_2$ ) material ($1200^oC$) sharply (to $600^oC$) 
because of a 
loss in global connectivity as some $Q^3$ units ($m= 1$) emerge\cite{5} 
at the expense 
of  $Q^4$ ($m= 0$) ones. The sharp reduction of $T_g$ destroys the mechanical 
equilibrium that prevailed\cite{5,6} in the pristine glass ($x = 0$), 
and drives 
alloyed glasses to become stressed rigid. This is largely the case because 
the bond-bending constraint of bridging oxygen atoms that were intrinsically 
broken\cite{6,7} at $1200^oC$, becomes restored in the weakly alloyed glass 
as $T_gs$ plummet to $600^oC$. However, upon continued addition of $Na_2O$  
the alloyed glass 
softens as network connectedness decreases, and one
expects an elastic phase transition from a 
stressed rigid to a floppy phase.
\par
One can estimate the elastic phase boundary within a mean-field theory by 
counting the Lagrangian constraints\cite{8,9} per atom ($n_c$) due to  
bond-stretching 
and bond-bending forces. The floppy mode fraction, $f (x) = (3-n_c) /3$ 
extrapolates linearly\cite{8,9} to zero as $n_c$ increases to $3$ 
defining the phase 
boundary. Physically, $f (x)$ represents the count of zero-frequency (floppy 
mode) solutions of the dynamical matrix in studying normal modes of a 
network. In the present oxide glasses, we  will show latter that the 
condition $n_c = 3$ is met\cite{7}  when  $x = 1/5$, and glass compositions 
at $x > 1/5$ 
are viewed as floppy while those at $x < 1/5$ stressed rigid. In the 
stressed-rigid phase, numerical simulations\cite{10,11} on random 
networks using 
a Kirkwood-Keating potential  have shown that  both longitudinal ($C_{11}$) 
and  shear ($C_{44}$) elastic constants  display a power-law variation as a 
function of $\bar r$ (mean coordination number) or $x$ , i.e., 
$C_{11}$  or $C_{44}\simeq (\bar r - \bar r_c)^p$  with $p = 1.5 (1)$ 
with a pronounced accuracy at higher $\bar r$. Such a power-law has been 
observed 
in Raman optical elasticities\cite{12,13}  but not in bulk 
elasticity\cite{14,15,16} measurements.
\par
We have now examined the rigidity onset in $(Na_2O)_x(SiO_2)_{1-x}$ 
glasses using  Brillouin scattering (BS) and Temperature Modulated 
Differential Scanning Calorimetry (MDSC). In this Letter, we show that the 
optical method\cite{15} reveals the mean-field behavior of the elastic phase 
transition: the elastic energy change $\Delta \Phi(x)$ upon thermal annealing
of the as 
quenched (virgin) glasses is lowered linearly 
at $x > x_c(2) = 0.238$, but to nearly vanish ($\simeq0$) at $x < x_c(1)= 0.18$. 
The observed variation in $\Delta \Phi(x)$ closely parallels the floppy mode 
fraction $f(x)$ in random networks\cite{10}, and serves to uniquely 
fix glasses at $x > 0.238$ to be floppy while those at $x < 0.18$ to 
be stressed rigid.
The stressed-rigid nature of the glasses at $x < 0.18$ is confirmed in 
Brillouin longitudinal ($C_{11}(x)$) and shear ($C_{44}(x)$) elastic constants 
that show a power-law variation with a power $p$ of $1.68 (8)$ and $1.69(8)$ 
in remarkable agreement with numerical simulations\cite{10,11}.  The 
non-mean-field behavior of the underlying rigidity transition is, 
however, manifested in MDSC that probes glasses at all length scales. 
In these calorimetric measurements, the non-reversing enthalpy ($\Delta H_{nr}$)  
at the glass transition ($T_g$) is found to be large at $x > x_c(2) = 0.238$ 
and at $x < x_c(1) = 0.18$, but to nearly vanish in the $x_c(1) < x < x_c(2)$ 
range. The latter range, the thermally reversing window\cite{17}, is 
identified with intermediate phases. Thus, the use of two complimentary
 probes, a mean-field (optical) and a non-mean-field (calorimetric) one, 
 has provided a rather comprehensive view of the three elastic phases, 
floppy, intermediate and stressed-rigid populated in an oxide glass system. 

$18$ gram $(Na_2O)_x(SiO_2)_{1-x}$ glasses over the soda 
concentration range $0.05  < x < 0.30$,  were  synthesized by  reacting the 
starting materials in a Pt-Rh crucible heated 
electrically to $1500^oC- 1650^oC$ for up to 4 hours. Details of the 
facility are described elsewhere\cite{18}. Melts were 
poured on stainless plates. Glass sample cubes $8~mm$ across were prepared 
by cutting the ingots with a diamond saw and polishing the surfaces to 
an optical finish. Brillouin scattering was excited with $\lambda=514.5~nm$ 
radiation and recorded in a right angle geometry using a pressure 
scanned\cite{18}, triple-pass plane Fabry-Perot interferometer with 
an effective finesse of 70.
Spectra were recorded for virgin- and annealed-glasses. The latter 
samples were obtained by heating virgin glasses at $515^oC$ for $4$ hours. 
Because of the hygroscopic nature of samples, the glasses were handled in a
controlled environment using glove boxes. MDSC measurements were 
performed\cite{13,17} using a model 2920 unit from TA Instruments Inc., 
at a scan rate of $3^oC/min$ and a modulation rate of $1^oC/100s$ using 
Au pans. MDSC measurements were undertaken on both virgin and  
vacuum annealed ($80^oC$ for 48 hours) glasses. 
\par
Brillouin lineshapes observed in the glasses ( Fig. 1) show 
the Longitudinal Acoustic (LA) mode frequency, $\nu_{LA}(x)$, to shift to a lower 
frequency as $x$ increases from $0.137$ to $0.20$ , and then to a
higher frequency at $x = 0.32$. Furthermore, mode frequency shifts due to 
annealing (broken-line curves) are small at low $x$ ($0.137$) but pronounced 
at high $x$ ($0.32$). The  longitudinal ($C_{11}$) and  shear ($C_{44}$) elastic 
constants were obtained using the relation, $C_{11} =\rho\nu_{LA}^2\lambda^2/2n^2$  
and $C_{44} = \rho\nu_{TA}^2\lambda^2/2n^2$  where $\nu_{LA}$, $\nu_{TA}$, $n$  and $\rho$ 
represent the 
LA- and   TA- mode frequencies, refractive index and mass density 
respectively. The refractive index was measured using a differential 
path refractometer and mass densities of virgin and annealed glasses
by buoyancy method. Note that density change is here achieved by
chemical alloying which influences the Brillouin lineshapes, in contrast
with density changes induced by pressure or temperature \cite{Grimsditch}.
\par
Compositional trends in $C_{11}(x)$ and $C_{44}(x)$ appear 
in Fig. 2, and show $C_{11}$ to decrease with $x$ at first, and then to 
increase at $x > 0.20$, due to layered-like disilicate units\cite{19}
 (adamantine) 
emerging as $x$ increases to $1/3$. On the other hand, the shear elastic 
constant $C_{44}$ systematically decreases as $x$ increases to $1/3$.  
We have also obtained the elastic energy change, $\Delta\Phi(x)$, upon 
thermal annealing glasses using\cite{18}:
\begin{eqnarray}
\Delta\Phi={\frac {1}{6}}\biggl({\frac {\Delta\rho}{\rho}}\biggr)(3C_{11}-4C_{44})
\end{eqnarray}
and find $\Delta\Phi$ to increase linearly at $x > 0.238$, and to vanish  at 
$x < 0.18$ as shown in Fig.2b.  The free energy of the thermally
relaxed glass is lowered by an amount $\Delta\Phi$ in relation to the virgin
glass. Calorimetric results 
on the glasses using MDSC appear in Fig.3, with (a) displaying variations 
in $T_g(x)$ deduced from the reversing enthalpy (b) variations in the 
non-reversing enthalpy, $\Delta H_{nr}(x)$.  We find $\Delta H_{nr}$ term is large 
at $x > 0.238$ and at $x < 0.18$, but the term almost vanishes ($~0$) in the 
$0.18 < x < 0.23$ range particularly in the annealed glasses. 
\par
Our interpretation of the BS results is as follows. Vibrational 
analysis of random networks within a mean-field theory reveals the floppy 
mode fraction as a function of $\bar r$ to be given\cite{10} by:
$f(\bar r) = 6-5\bar r/2$.
 In the present glasses, since $\bar r = (8-4x)/3$ (ref. \cite{21}), 
the latter equation can be written as: $f(x) = {\frac {10}{3}}x - 
{\frac {2}{3}}$ which reveals $f(x)$ to increase linearly at $x > 1/5$ in the 
floppy phase, and $f(x)\simeq 0$ at $x < 1/5$ in the stressed-rigid phase, 
thus, 
showing the mean-field elastic phase boundary at $x_{mf}  = 1/5$.  More accurate 
numerical simulations based on a bond-depleted amorphous Si network 
performed by M.F.Thorpe\cite{10} have localized the phase boundary by plotting 
the second derivative of $f(\bar r)$, which shows a 
cusp at $\bar r = 2.385$. 
For the case of the present oxides, the corresponding phase boundary would 
be at\cite{21}  $x_{num} =  0.211$. The observed variation in $\Delta\Phi(x)$
 (Fig.2b) mimics 
the results of these numerical simulations of $f(x)$ , and $d^2\Delta\Phi(x)/dx^2$ 
shows 
a cusp near $x = 0.238$.  Quenched glasses at $x > 0.238$
relax\cite{18} as frozen stress is thermally annealed away by 
floppy or bond-rotational modes. These modes are associated 
with the underconstrained $Q^2$ and $Q^1$ units that grow\cite{5} 
precipitously 
at $x > 0.25$. These modes are nearly absent in the more connected glasses 
($x < 0.18$).  Furthermore, since our glasses were relaxed at a fixed 
annealing 
temperature of $515^oC$, $\Delta\Phi(x)$ provides a measure of elastic free energy in 
much the same fashion that $f(x)$ is viewed\cite{10} as the network 
free energy.  
The correlation serves to uniquely fix glasses at $x > 0.238$ to be floppy 
while those at $x < 0.18$ to be stressed-rigid. Here we must recall that 
the length scale over which BS probes the elastic behavior of glasses is 
set by the wavelength of the acoustic phonons that lie in $\lambda\simeq 300~nm$ 
range\cite{Courtens}.  
One, therefore, expects BS to probe the average elastic behavior of the 
rigidity transition in the present oxide glasses as in 
chalcogenide glasses\cite{15}. 
\par 
The calorimetric probe (MDSC) registers enthalpic changes near $T_g$ in a glass 
network, and these contributions come from molecular rearrangements taking 
place at all length scales. And for that reason, one expects to observe 
non-mean-field effects associated with the elastic phase transition using 
the thermal probe. The existence of a thermally reversing window (Fig. 3b) 
is an example of such an effect\cite{17}.  We find the window  
sharpens and deepens 
upon low-temperature thermal anneal because network stress frozen upon a 
quench is relieved. The window in the present oxide glasses is reminiscent 
of parallel results in chalcogenide glasses\cite{12,13,17}. The observation of 
a reversibility window in the present oxides leads naturally to the 
suggestion that glasses in the $0.18 < x < 0.238$ composition range are in 
the intermediate phase, i.e., they are self-organized. In this range of 
composition,  $Si^{29}$  NMR shows\cite{5} concentration of  
undersconstrained\cite{21} $Q^3$ 
($n_c = 2.67$ )  and overconstrained $Q^4$  ($n_c = 3.67$)  
units to be comparable. 
Self-organization effects stem\cite{23} from the fact such a mix leads to an 
isostatically rigid global structure ($\bar n_c\simeq 3$).
\par
The stressed rigid nature of glasses at $x < 0.18$ is confirmed by the 
power-law variation in $C_{11}(x)$ and $C_{44}(x)$. The variation is inferred by 
plotting $log(C_{11}(x) - C_{11}(x_c(1))$ against $log (x_c(1)- x)$ , 
using $x_c(1) = 0.18$, and yields (Fig. 4) $p=p_{11} =1.68 (8)$. Parallel 
analysis of $C_{44}(x)$  
yields a power-law $p = p_{44} = 1.69(8)$. These results are in remarkable 
agreement with numerical predictions\cite{10,11} of $p = 1.5 (1)$ in random 
networks as noted earlier in Raman scattering on chalcogenide 
glasses\cite{12,13}. 
To the best of our knowledge this is the first time one has observed a 
power-law variation in elastic constants for stressed rigid glasses 
using a bulk probe. In BS results on chalcogenide glasses\cite{15}  as in 
ultrasonic elastic moduli results\cite{14,16} one could discern only a linear 
variation in $C_{11}(x)$. Why is the rigidity transition washed out in 
chalcogenide- but not in the present oxide glasses ?  Residual interactions 
due to dihedral angle forces and lone-pair van der Waals forces 
dilute the effect of  first-neighbor (bond-stretching) and second-neighbor 
(bond-bending) forces and wash out the phase transition. We note that the 
$Si-O$ single bond strength\cite{24} ($100~kcal/mole$) far exceeds the $Ge-Se$ 
single bond strength ($40~kcal/mole$). Furthermore, lone pair
interactions
are likely to be much weaker in the oxides
than in chalcogenide glasses. 
It appears that the power-law variation in elastic 
constants is observed in those instances where first- and second- 
neighbor forces overwhelm residual interactions as is the case of 
Brillouin scattering in the oxides, or that of  Raman scattering in 
chalcogenide\cite{12,13}. 
\par
In summary, Brillouin scattering and MDSC have permitted identifying the
three elastic phases in $(Na_2O)_x(SiO_2)_{1-x}$ glasses; glasses 
at $x < 0.18$ are stressed-rigid, those in $0.18 < x < 0.23$ are
intermediate, while those at $x > 0.23$ are floppy. 
A power-law variation of the longitudinal- and shear- elastic constants 
is observed in the stressed-rigid phase of the oxide glasses using a 
bulk probe (BS) for the first time. 
These novel results show that the floppy to stressed-rigid phase
transitions in oxide and chalcogenide glasses are remarkably similar,
underscoring the basic physics driving formation of elastic phases
in disordered systems. Our results correlate well with electrical
transport that show activation energy for diffusion \cite{Greaves} in
the floppy glasses to be consistently low, and to increase monotically
as the backbone becomes increasingly stressed-rigid. 
This work is supported by a joint CNRS-NSF collaboration research 
project 13049 and NSF grant Int- 01-38707 and DMR-01-01808.           .       

\newpage
\listoffigures
\newpage
\begin{figure}
\begin{center}
\epsfig{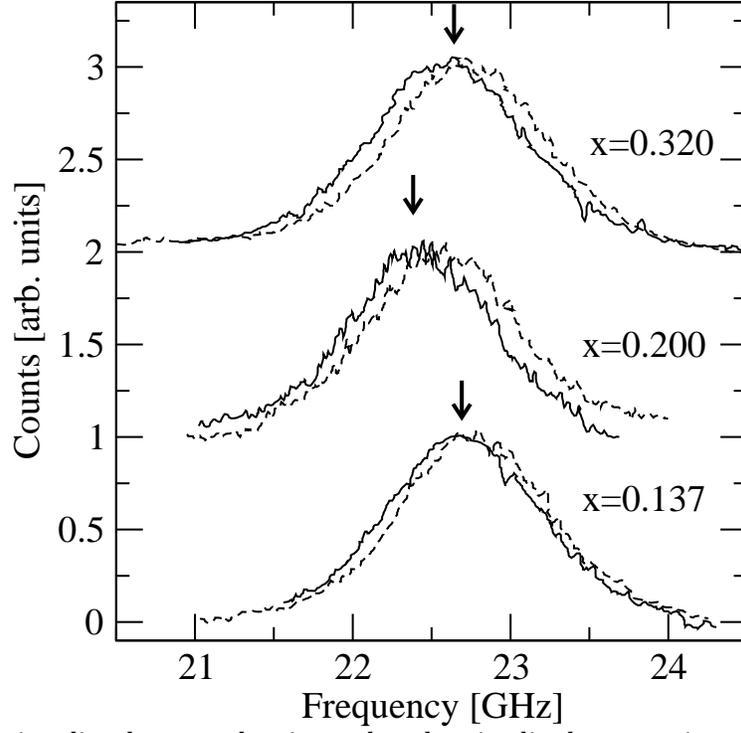}
\caption{Brillouin lineshapes showing the longitudinal acoustic mode at 
indicated
$(Na_2O)_x(SiO_2)_{1-x}$ glass compositions $x$. The mode shift between virgin 
(continuous line) and annealed (broken line) glasses is small at low $x$ 
($0.137$) and large at high $x$ ($0.32$).}
\end{center}
\end{figure}
\newpage
\begin{figure}
\begin{center}
\epsfig{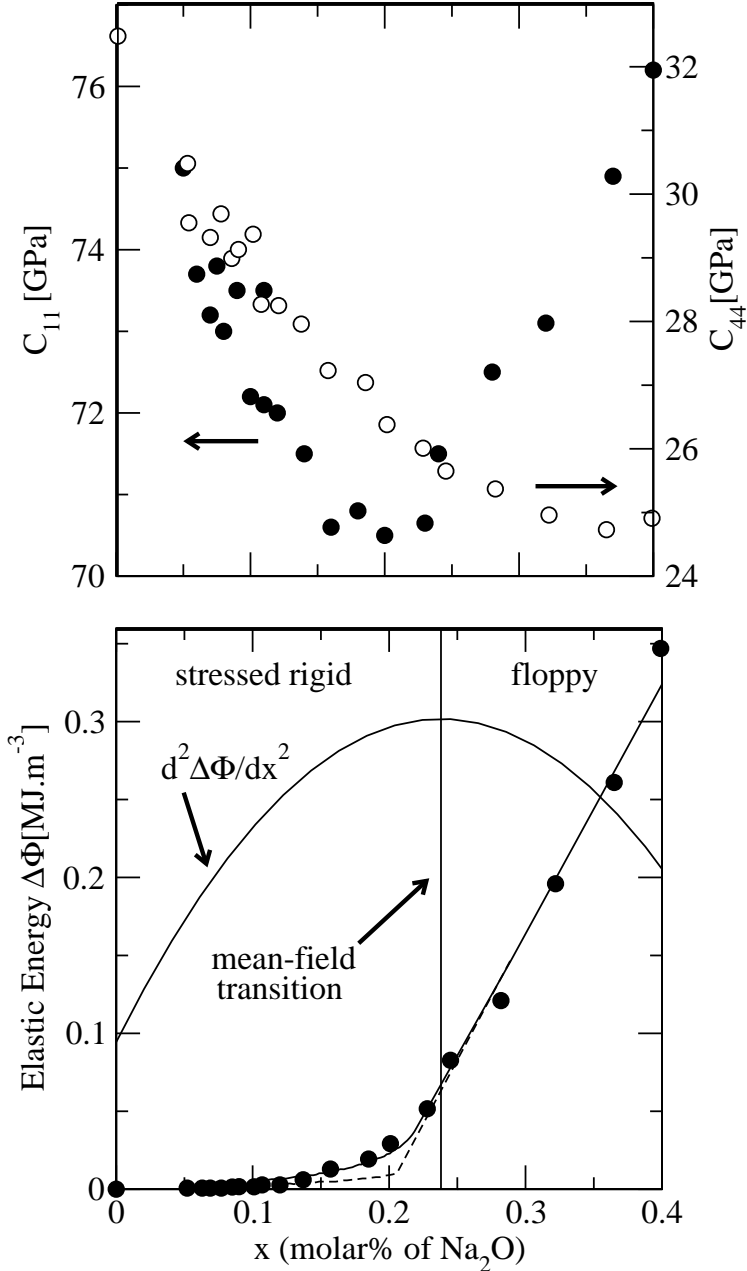}
\caption{(a) Variations in $C_{11}(x)$ and $C_{44}(x)$ and  (b)  in elastic 
energy increase $\Delta\Phi(x)$ in $(Na_2O)_x(SiO_2)_{1-x}$ glasses deduced from 
Brillouin scattering. The continuous line and broken line give the floppy 
mode fraction $f(x)$ prediction for  random- and self organized networks 
respectively. The second derivative of $\Delta\Phi(x)$ obtained from the BS 
results fix the mean-field elastic phase transition at $x = 0.238$, 
with glasses at 
$x > 0.238$ floppy.}
\end{center}
\end{figure}
\newpage
\begin{figure}
\begin{center}
\epsfig{figure=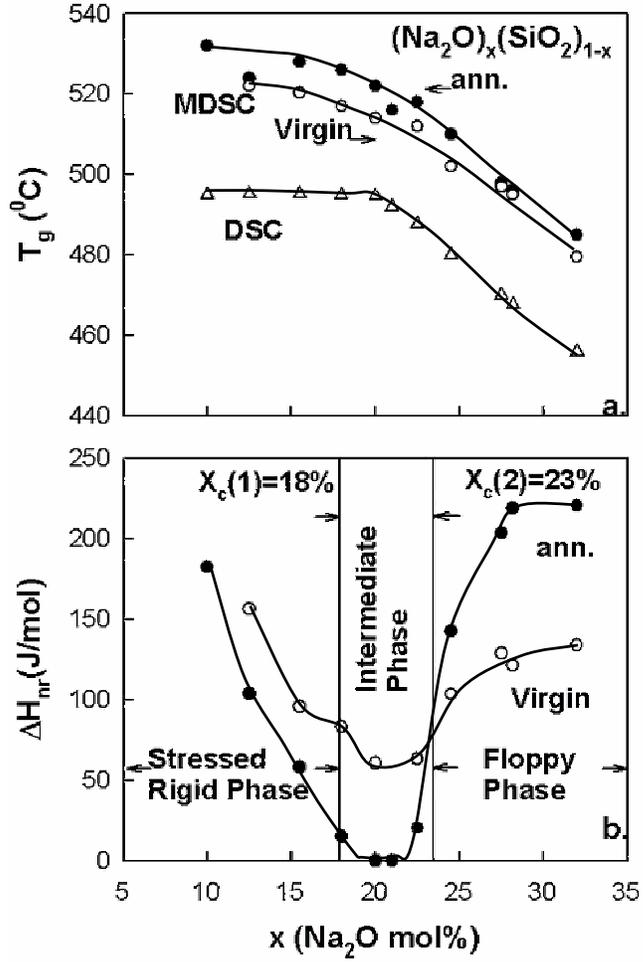,width=0.8\linewidth}
\caption{MDSC results on $(Na_2O)_x(SiO_2)_{1-x}$ glasses showing variations 
in (a) $T_g(x)$ , (b)
non-reversing enthalpy $\Delta H_{nr}(x)$ for virgin ($\circ$) and annealed 
($\bullet$) samples. The DSC results were taken from ref. 26.}
\end{center}
\end{figure}
\newpage
\begin{figure}
\begin{center}
\epsfig{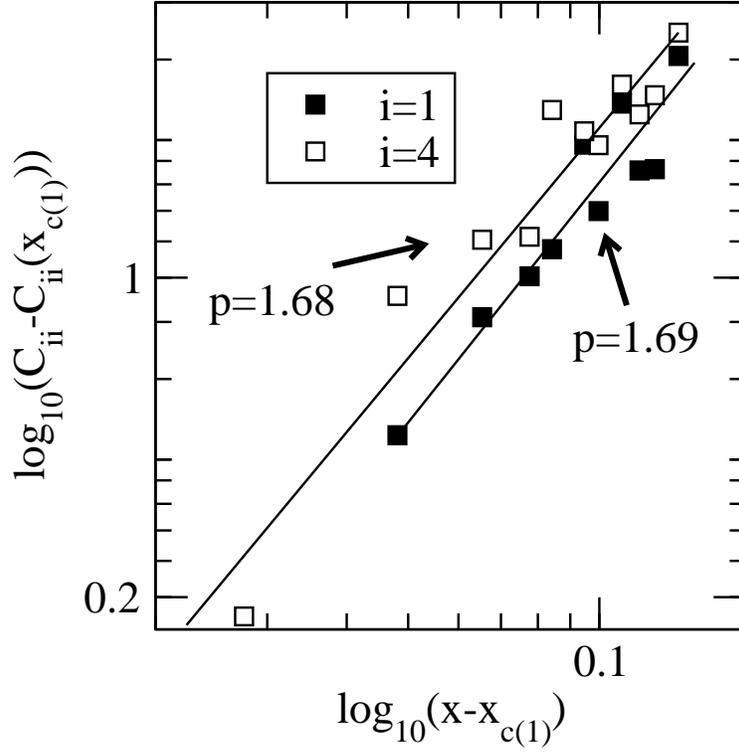}
\caption{A log log plot of $C_{11}(x)-$ $C_{11}(0.18)$ against $( x - 0.18)$ 
showing a power-law variation  with a power $p = 1.69(8)$ in the 
stressed-rigid region ($0 < x < 0.18= x_c(1)$). Corresponding results 
for $C_{44}(x)$ show a power-law of $1.68(8)$.See text for details. }
\end{center}
\end{figure}

\begin{thebibliography}{100}
\bibitem{21} In  $(Na_2O)_x(SiO_2)_{1-x}$ glasses, taking the coordination  
number of Si, Na and O to be respectively 4, 1 and 2, we obtain  
$\bar r = [4x +8(1-x)]/3 =  (8-4x)/3$; For a  $\bar r = 2.385$, the 
implied threshold in $x_{num} = (8 - 3r)/4 = 0.211$ for the present oxides; 
For $Q^4$ [$Si(O_{1/2})_4$] units, the Lagrangian constraints/ atom, 
$n_c = (1/3)[ 7 + 2×{\frac {1}{2}}× 4] = 3.67$; For $Q^3$ 
[$Si(O_{1/2})_3O^-Na^+$] units, $n_c = (1/4.5)[ 7 + 3× {\frac {1}{2}}× 
2 +1 +1] = 2.67$.
\end{thebibliography}
\end{document}